\documentclass[prl,twocolumn,showpacs,]{revtex4}
\usepackage{ae}
\usepackage{graphicx}
\usepackage{amssymb}
\usepackage{amsmath}
\usepackage{amsthm}
\usepackage{color}
\usepackage{sidecap}
\usepackage{ulem}
\usepackage{wasysym}
\usepackage{pifont}
\usepackage{ifsym}
\def\nosop{\hat{n}_{\rm OS}}
\def\nosopone{\hat{n}_{\rm OS 1}}
\def\nosoptwo{\hat{n}_{\rm OS 2}}

\def\Jperp{J_{\perp}}
\def\Iem{\mathcal{I}^{\rm em}}
\def\Iab{\mathcal{I}^{\rm ab}}

\begin{document}
\title{Lattice assisted spectroscopy: a generalized scanning tunnelling microscope for ultra-cold atoms.}

\author{A.~Kantian$^1$, U.~Schollw\"ock$^2$, T.~Giamarchi$^3$}
\affiliation{(1) Nordita, KTH Royal Institute of Technology and Stockholm University, Roslagstullsbacken 23, SE-106 91 Stockholm Sweden\\
  (2) Department f\"ur Physik and Arnold-Sommerfeld-Centre for Theoretical Physics, LMU M\"unchen, Theresienstrasse 37, 80333 M\"unchen, Germany\\
  (3) DQMP, University of Geneva, 24 Quai Ernest-Ansermet, 1211 Geneva, Switzerland }

\begin{abstract} 
We show that the possibility to address and image single sites of an optical lattice, now an experimental
reality, allows to measure the frequency-resolved local particle and hole spectra of a
wide variety of one- and two-dimensional systems of lattice-confined strongly correlated ultracold atoms. Combining
perturbation theory and time-dependent DMRG, we validate this scheme of lattice-assisted spectroscopy (LAS) on several example
systems, such as the 1D superfluid and Mott insulator, with and without a parabolic trap, and finally
on edge states of the bosonic Su-Schrieffer-Heeger model. We highlight important
extensions of our basic scheme to obtain an even wider variety of interesting and important frequency resolved
spectra.
\end{abstract}
\date{\today}
\maketitle
Probing time- or frequency-dependent response functions of strongly correlated systems
forms the practical foundation for much of the current study of condensed matter. Among these, single particle observables have a special role,
often constituting \textit{the} basic measurements for determining a materials properties. Even individual techniques
in this field (STM, ARPES) form entire subbranches of experimental and associated theoretical physics of their own~\cite{Damascelli2003, Fischer2007}.
To have such fundamentally important techniques also available for ultracold atomic gases has been the subject
of intense research activity. For fermionic atoms, initial work focussed on gases in parabolic traps, first on obtaining the gap
for paired fermions~\cite{Torma2000} by frequency-resolved measurements, then expanded to obtain the fully
momentum-resolved spectral function via Bragg spectroscopy~\cite{Dao2007,Stewart2008}.
Performing this on systems that only experience an overall confining potential but no additional
optical lattice is due to a simple reason: typically, numbers in lattice-confined systems are too small to obtain a significant
signal. Also, such global measurements in presence of a trap come at a price, as the obtained Greens function may have their momentum-dependencies modified~\cite{Frohlich2012}.
\begin{figure}
\begin{center}
\vspace{-0cm}
\includegraphics[width=1\columnwidth,trim = 59mm 27mm 70mm 27mm, clip]{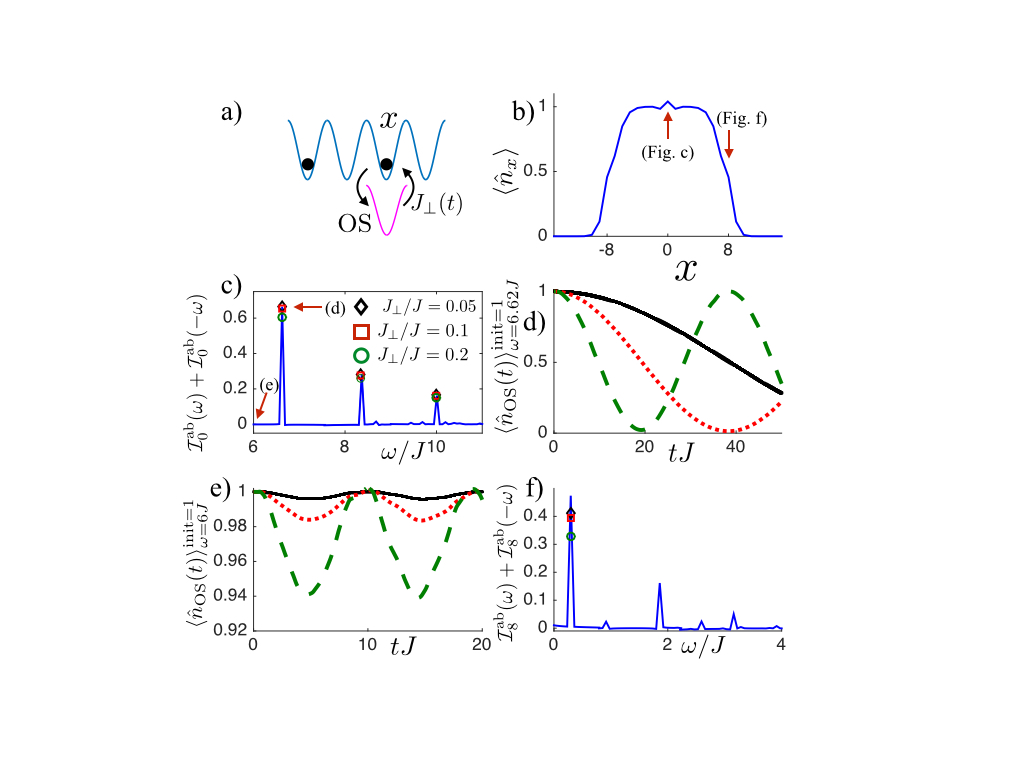}
\end{center}
\caption{(Color online) {\bf a)} Schematic overview of spectroscopic scheme: an extra outcoupling `site' (OS) (physical or internal state) is tunnel-coupled
with amplitude $\Jperp(t)$ to lattice site $x$. The time-evolving average $\langle \nosop(t)\rangle$ on OS is obtained from repeated occupation-number imaging. {\bf b)} Local density for system in test case {\bf (A)}, $15$ bosons in a 1D parabolic trap. The MI region is in the centre, and SF regions occupy the wings. Red arrows indicate sites where system is locally probed. {\bf c)} Symmetrized discrete absorption spectrum in centre of MI region, extracted from the full time-evolving average $\langle\nosop(t)\rangle_\omega^{\rm init=1}$ for $\Jperp/J=0.05$ (black $\Diamond$), $=0.1$ (red $\square$), $=0.2$ (green $\circ$), and compared with $\mathcal{I}^{\rm ab}_x(\omega)+\mathcal{I}^{\rm ab}_x(-\omega)$ obtained directly from DMRG (blue solid line). {\bf d)} $\langle\nosop(t)\rangle_\omega^{\rm init=1}$ over time at first resonance $\omega=6.62J$ for $\Jperp/J=0.05$ (black solid line), $=0.1$ (red dotted), $=0.2$ (green dashed). Rabi-oscillations can be clearly observed at experimentally accessible timescales. {\bf e)} $\langle\nosop(t)\rangle_\omega^{\rm init=1}$ over time below first resonance, at $\omega=6J$, vales of $\Jperp/J$ match d). Off-resonance, oscillations lack all hallmarks of Rabi dynamics, exhibiting small amplitudes scaling with $\Jperp/J$, while frequency does not depend on it. {\bf f)} Equivalent figure to c) for site $8$ in the SF region.}   
\label{fig:Fig1}
\end{figure}

Yet, the presence of an optical lattice is key for the many proposed uses of ultracold atomic gases
to serve as quantum simulators of strong correlation physics~\cite{Bloch2008}, and practical ways for measuring time/frequency-resolved
observables in such systems have to be developed. Applying Bragg spectroscopy,
as has been done for 1D bosonic Mott-insulators (MI), sidesteps any potential problems with insufficient atom numbers by performing
the measurements on 2D arrays of decoupled 1D chains, to obtain the spectrum of density-density~\cite{Stoferle2004}
or single-particle excitations~\cite{Fabbri2012}. Performing Bragg spectroscopy in parallel in this way would however
not be practical for 2D systems, and further presupposes that the physics of each 1D chain is not too affected by the
inevitable density variations across the 2D array of decoupled chains. An alternative probing proposal, based on using a confined ion as a 'tip'~\cite{Kollath2007} is still in practical infancy (though the setup has been partially realised~\cite{Zipkes2010})

Critically, the Bragg approach will fail for the latest generation of quantum gas
microscope experiments~\cite{Bakr2009,Sherson2010,Simon2011,Cheneau2012,Fukuhara2013,Haller2015}. These may offer both single-site resolution and addressing, but only across one or two parallel 2D planes, for what will essentially
be systems of a few thousand atoms at most. Yet, the future study of strongly correlated atomic gases is moving very much in this direction, with multiple groups pushing to build up quantum gas microscopes. On the other hand, single-site addressing offers a very different alternative approach to measure frequency-resolved single particle
excitations. Some of us~\cite{Knap2013} co-authored a recent proposal to leverage these new experimental capabilities to measure arbitrary
time-dependent spin-spin correlators for any Hamiltonian implemented in an optical lattice that maps to an effective spin Hamiltonian.

In this work, we show how to use single-site resolution and addressing to obtain
time-dependent local particle and hole Greens functions for \textit{any} Hamiltonian that can be prepared in a quantum gas
microscope setup. We call this scheme \textit{lattice-assisted spectroscopy} (LAS) in the following. Using time-dependent density matrix renormalization group (t-DMRG)~\cite{Schollwock2011} simulations we quantitatively validate the LAS-scheme on both superfluid (SF) and MI systems, as well as on topologically protected edge states of the bosonic
Su-Schrieffer-Heeger (SSH) model. We conclude with highlighting some of the most relevant of the many possible extensions of LAS.

We consider the situation shown in Fig.~\ref{fig:Fig1} a: the system, assumed to be in the equilibrium state
of some lattice Hamiltonian $\hat{H}$ (for which we want to measure
local Greens functions), as well as an `extra' site, denoted as the \textit{outcoupling `site'} (OS) in the following. The OS could denote another internal state of the atoms into which they are weakly coupled only on site $x$, exploiting site-dependent addressing. Experiments have already demonstrated this capability~\cite{Fukuhara2013} - thus all required techniques to implement LAS already exist. Alternatively, the OS could be another physical site
initially separate from the system, with controllable tunnel-coupling to system-site $x$. Such a setup should be readily realisable with existing techniques - digital micromirror arrays integrated into quantum gas microscopes have already been used to write a large variety of spatially varying potentials resolved on the single-site scale~\cite{Zupancic2013}. With this, an OS could be created by allowing weak tunnelling from a system site only onto a single site in an empty parallel lattice, by imprinting a large energy-offset on every other site of the empty lattice.

In either case, we assume an oscillating  tunnel-coupling between atoms on site $x$ and atoms in the OS. Its time-dependency may either be of form $\Jperp(t)=\Jperp\cos(\omega t)$ or $\Jperp(t)=\Jperp\sin(\omega t)$, switched on at time $t=0$. 
Now, the resultant time-evolution of the average occupation on the OS will depend on the initial state of the OS. In the following, we focus on the most practically relevant cases, where there is initially either no atom on the OS, 
\begin{equation}\label{lin_resp_eq_init_empty}\langle \nosop (t) \rangle_\omega^{\rm init=0} = \frac{\Jperp^2}{\hbar^2}\int d\omega' \tilde{\mathcal{K}}_t(\omega,\omega')\Iem_x(\omega'-\delta h_{\rm}(0,1))\end{equation}
or exactly one: 
\begin{equation}\label{lin_resp_eq_init_occ}
\begin{split}
\langle \nosop (t) \rangle_\omega^{\rm init=1} & = 1 + \frac{\Jperp^2}{\hbar^2}\int d\omega' \tilde{\mathcal{K}}_t(\omega,\omega')\left[\right.  2\Iem_x(\omega'-\delta h_{\rm}(1,2)) \\
  & - \Iab_x(\omega'-\delta h_{\rm}(1,0)) \left. \right].
\end{split}
\end{equation}
In eq. (\ref{lin_resp_eq_init_occ}), the first term is only present for bosons and absent for fermions. We also introduced the shorthand $\langle \nosop (t) \rangle_\omega=\langle \nosop (t) \rangle_{\cos \omega}+\langle \nosop (t) \rangle_{\sin \omega}$. We assume that site OS has internal dynamics described
by the Hamiltonian $\hat{h}_{\rm OS}=h_{\rm OS}(n)$, diagonal in the occupation number $n$, with $\delta h_{\rm OS}(n,m) := h_{\rm OS}(n)-h_{\rm OS}(m)$. It is also assumed that only a single species of atoms may tunnel between sites $x$ and OS, but LAS-formulas for multi-component systems are straightforward to derive.

Here, the time-evolving averages are computed by considering tunnelling as a quadratic perturbation,
which links them to the local \textit{emission [absorption] spectrum} of the system:
\begin{equation} 
\mathcal{I}^{\rm em [ab]}_x(\omega') = \sum_{n,m} |\langle n |\hat{a}_x [\hat{a}^{\dagger}_x] |m \rangle |^2 \frac{e^{-\beta E_m}}{\mathcal{Z}}\delta(E_m-E_n-\omega'),
\end{equation}
integrated over with the kernel
\begin{equation}
\tilde{\mathcal{K}}_t(\omega,\omega')= 2\sum_{\sigma=\pm}\frac{\left( 1 - \cos((\omega'+\sigma\omega)t) \right)}{(\omega'+\sigma\omega)^2},
\end{equation}
where $|n\rangle$, $E_n$, $\mathcal{Z}$ denote the exact eigenvectors, eigenenergies and partition function of the system Hamiltonian $\hat{H}$.
This kernel has the known property $\tilde{\mathcal{K}}_t(\omega,\omega')/t \stackrel{t\rightarrow\infty}{\rightarrow}\delta(\omega-\omega')+\delta(\omega+\omega')$. If the probing time is not effectively infinite w.r.t. all other time-scales and the spectrum is not dense however, the implications of the kernel properties for eqs.~(\ref{lin_resp_eq_init_empty}),~(\ref{lin_resp_eq_init_occ}) fall broadly into one of
two possible regimes: {\it the excited states in $\mathcal{I}^{\rm em[ab]}_x(\omega')$ appear dense to $\mathcal{K}_t(\omega,\omega')$ at probing frequency $\omega$ and time $t$}, or alternatively {\it only one state lies close or at frequency $\omega$ inside the kernel envelope}. In the first regime, the occupation on OS will change {\it linearly}, $\langle \nosop (t) \rangle_\omega \rightarrow \frac{\Jperp^2}{\hbar^2}\left(\mathcal{I}_x(\omega) + \mathcal{I}_x(-\omega)\right)t$ - thus, the symmetrized $\mathcal{I}_x(\omega)$ can be extracted from a linear fit. In the second regime, at short times when the kernel is still very broad, $\langle \nosop (t) \rangle_\omega$ will also behave linearly, as other eigenstates may still contribute. For longer times however,
when $\tilde{\mathcal{K}}_t(\omega,\omega')$ narrows enough to filter out the discrete level, $\langle \nosop (t) \rangle_\omega \rightarrow \frac{\Jperp^2}{\hbar^2}
\left(\sum_{n,m: E_m-E_n=\pm\omega}|\langle n | \hat{O} | m \rangle |^2 \frac{e^{-\beta E_m}}{\mathcal{Z}}\right)t^2$; here $\hat{O}=\hat{a}_x [\hat{a}^{\dagger}_x]$.  At even longer times, as the population of
atoms on the OS begins saturating to the physically permitted level, we expect (and observe; see below) $\langle \nosop (t) \rangle_\omega$ to begin deviating from
the perturbative prediction strongly. On this largest timescale in the second regime, $\langle \nosop (t) \rangle_\omega$ performs what are essentially Rabi oscillations, with
frequency equal to  $\frac{\Jperp}{\hbar}
\sqrt{\sum_{n,m: E_m-E_n=\pm\omega}|\langle n | \hat{O} | m \rangle |^2 \frac{e^{-\beta E_m}}{\mathcal{Z}}}$ at small enough $\Jperp$, while the perturbative expression will naturally keep behaving monotonically. As the range of validity in time $t$ and/or $\Jperp$ is a key issue with any formulas from time-dependent perturbation theory we use t-DMRG in the following to validate LAS, and specifically eqs.~(\ref{lin_resp_eq_init_empty}) and (\ref{lin_resp_eq_init_occ}), by comparing them to the full evolution of $\langle \nosop (t) \rangle_\omega$. (where $\mathcal{I}_x(\omega)$ is determined independently, here also via DMRG).

When experimentally applying LAS, both of the above regimes can be relevant. For one, cold atom experiments offering single-site addressability
currently involve a few tens of sites and particles. At sufficient frequency resolution (i.e. sufficiently long probing time) their bulk spectra will reveal their discrete nature. The other possibility (applicable even when future experiments
are so large that bulk spectra always appear dense), are localized gap-protected states, such as (topologically) protected edge states or states around some impurity. As a model Hamiltonian to study the validity of eqs.~(\ref{lin_resp_eq_init_empty}), (\ref{lin_resp_eq_init_occ}) for all these possibilities we use a generalised 1D Bose-Hubbard model:
\begin{align}\label{BHham}
\hat{H} & =  -\sum_{x=1}^{L-1} (J+(-1)^x\delta J)(\hat{a}^\dagger_x \hat{a}_{x+1}+{\rm h.c.} ) \nonumber \\
+ & \frac{U}{2}\sum_{x=1}^L \hat{n}_x(\hat{n}_x-1) + V_{\rm p}a^2\sum_{x=1}^L (x-(L+1)/2)^2 \hat{n}_x,
\end{align}
where $\hat{a}^\dagger_x$ creates a bosonic atom on site $x$ of an $L$-site chain with lattice constant $a$ (with open boundary conditions), tunneling between sites $x$ and $x+1$ takes place with amplitude $J+(-1)^x\delta J$, $V_{\rm p}$ denotes the strength of any parabolic trapping potential that might be applied, and $U>0$ denotes the strength of the onsite repulsion between atoms. 

We treated three test cases to study and demonstrate the capabilities of LAS. {\bf Test (A):} Bosonic atoms in a parabolic trap form a so-called 'wedding cake' when repulsive interactions are sufficiently strong~\cite{Bloch2008} (c.f. Fig.~\ref{fig:Fig1}). For $\delta J = 0$, $V_{\rm p} > 0$, $U\gg J$ MI areas will alternate with SF transition regions from the traps centre outward. The LAS-scheme allows probing of the gapped or non-gapped nature of these respective regions directly. We choose a system of direct relevance to current experimental capabilities, 15 bosons in a trap with $V_{\rm p}/J=0.06$, $U=10J$. Its absorption spectrum $\mathcal{I}^{\rm ab}$ can be obtained with great precision from quadratic fits to $\langle\nosop(t)\rangle^{\rm init=1}_\omega$ at short times together with observation of the Rabi oscillations at longer times (still occurring on experimentally accessible timescales), as outlined above, and summarised in Figs.~\ref{fig:Fig1} c-f. This directly reveals the size of the gap in the MI centre of the 'cake', and its disappearance in the wings. When the OS is coupled to a site in the MI and SF region respectively, we use the possible control over  $h(n) = V_{\rm p}a^2(x-(L+1)/2)^2 + U_{\rm OS}/2n(n-1)$ to shift the contribution of $\mathcal{I}^{\rm em}$ to the emission spectrum, eq.~(\ref{lin_resp_eq_init_occ}) to be effectively zero, where $U_{\rm OS}$ denotes the atom-atom interaction on OS. In the MI centre, this is done by having $U_{\rm OS}=0$, while in the SF wing this requires $U_{\rm OS}=U$. We stress that even for a quite nonperturbative $\Jperp=0.2 J$ only very modest signal broadening is observed.

\begin{figure}
\begin{center}
\vspace{-0cm}
\includegraphics[width=1\columnwidth,trim = 72mm 75mm 80mm 72mm, clip]{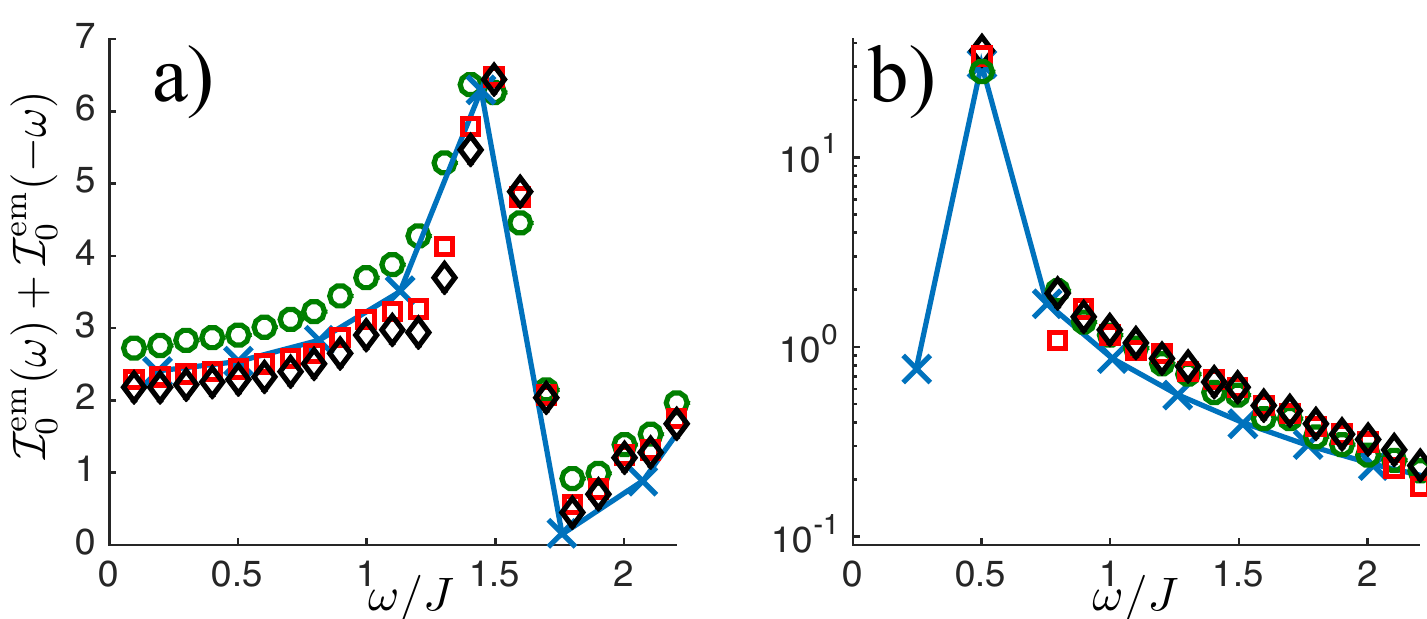}
\end{center}
\caption{(Color online) Finite-resolution emission spectra in the thermodynamic limit for MI and SF phases, reconstructed from $\langle\nosop(t)\rangle_\omega^{\rm init=0}$ for $\Jperp/J=0.05$ (black $\Diamond$), $=0.1$ (red $\square$), $=0.2$ (green $\circ$), and compared with $\mathcal{I}^{\rm em}_x(\omega)+\mathcal{I}^{\rm em}_x(-\omega)$ obtained directly from DMRG (blue $\times$, blue solid line is guide to the eye) {\bf a)} $U=10J$. {\bf b)} $U=2J$.}   
\label{fig:Fig2}
\end{figure}

For {\bf Test (B)}, we consider somewhat larger systems than previously, such as might be experimentally feasible in the near future. Here we want to show how LAS can be applied to obtain spectra that effectively approximate those of infinite systems, with a frequency resolution set by the finite size. For this, we focus on a Bose-Hubbard model with $L=N=51$, $\delta J = V_{\rm p} = h(n) = 0$ in both the MI and SF regimes. We obtainin the emission spectrum $\Iem_x(\omega)$ from $\langle\nosop(t)\rangle^{\rm init=0}_\omega$, from an OS coupling to the central site - note that we do not observe a gap in the MI regime as we are looking at the spectral function of a hole, and not of a particle (that would be $\Iab_x(\omega)$). For times $t$ smaller than those at which the excitations triggered by the probing hit the systems open boundaries, we find excellent agreement with the finite-resolution spectra obtained from the convolution $\int d\omega' \partial_t\tilde{\mathcal{K}}_t(\omega,\omega')\mathcal{I}^{\rm em}_x(\omega')$, which thus have a resolution set by the same time-scale (see Fig.~\ref{fig:Fig2}).

\begin{figure}
\begin{center}
\vspace{-0cm}
\includegraphics[width=1\columnwidth,trim = 62mm 92mm 73mm 95mm, clip]{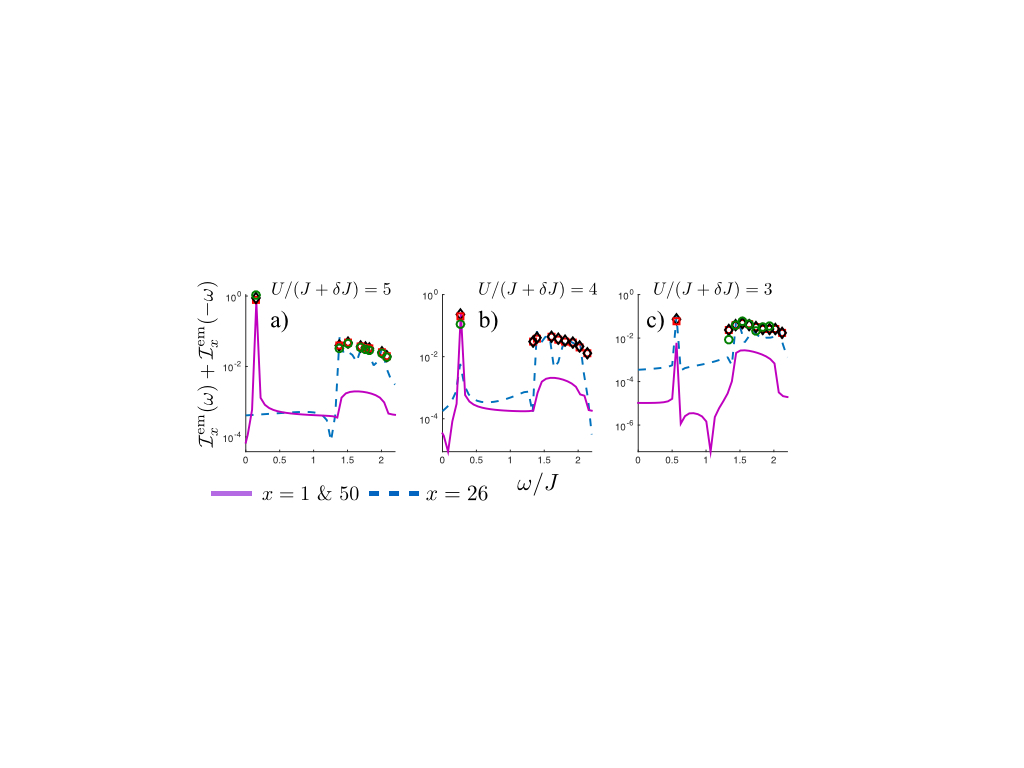}
\end{center}
\caption{(Color online) Bulk and edge spectra of bosonic SSH model with $L=50$. Markers denote spectra reconstructed from $\langle\nosop(t)\rangle_\omega^{\rm init=0}$ for three different $\Jperp/J$: $\Jperp/J=0.05$ (black $\Diamond$), $\Jperp/J=0.1$ (red $\square$) and $\Jperp/J=0.2$ (green $\circ$). For comparison, we show $\mathcal{I}^{\rm em}_x(\omega)+\mathcal{I}^{\rm em}_x(-\omega)$ obtained directly from DMRG (dashed blue lines for bulk at $x=26$ and solid purple lines for edge spectra, at $x=1$ and $=50$, respectively). While $U/(J+\delta J)=5$, the distinct edge mode persists, while for lower $U/(J+\delta J)$ bulk and edge spectra start becoming indistinguishable, with overall spectral weight on the edges collapsing.} 
\label{fig:Fig3}
\end{figure}

With {\bf Test (C)}, we highlight the power of LAS to probe the edge states present in topologically protected systems. For this, we study a system with 
$L = 2N = 50$, $\delta J = 2/3 J$, $V_{\rm p} = 0$, $h(n) = 0$. The bosonic version of the SSH model~\cite{Su1979} has edge states near zero energy while the bulk is gapped (as would be expected of a model like (\ref{BHham}) with finite dimerisation, $|\delta J|>0$) as long as the particle-hole gap of the system is finite~\cite{Grusdt2013}. As shown in Fig.~\ref{fig:Fig3}, LAS applied to both edges and bulk allows accurate observation of the distinct excitation spectra of both edges and bulk for larger $U$, and the gradual collapse of the distinct edge modes once $U\leq 4(J+\delta J)$.

Our quantitative DMRG results also put to rest a potential key issue for LAS, the degree to which changes in the magnitude $\langle\nosop(t)\rangle_\omega$ can be resolved in practice. With differences on the order of a few percent being readily detectable~\cite{Fukuhara2013}, in every test case the largest $\Jperp$ value we ran, $\Jperp=0.2J$, always yielded clear signal wherever there was appreciable spectral weight, and in most cases $\Jperp=0.1J$ gave useful data too. Even at $\Jperp=0.2J$, spectral weight can be too small though - this is why we do not compare spectra from $\langle\nosop(t)\rangle_\omega$ against the low-magnitude parts of $\mathcal{I}_x(\omega)$ in Fig.~\ref{fig:Fig3}. As we show in Figs.~\ref{fig:Fig1} - \ref{fig:Fig3} the signal broadening at these $\Jperp$ is modest - only for discrete excited state such as in the SSH model we find that amplitudes can be up to halved at $\Jperp=0.2J$ when the spectral weight is low (yet still detectable).

So far we have proposed and validated a general scheme to extract frequency-resolved local Greens functions for any system of interacting atoms confined to an optical lattice. Now we highlight several extensions to deal with nonlocal and higher-order correlations. First, if a single atom is created in a known superposition $(\alpha \hat{a}^{\dagger}_{\rm OS 1}+\beta \hat{a}^{\dagger}_{\rm OS 2})|\emptyset\rangle$ 
between \textit{two} OS (e.g. by having a single initially single-site localized atom delocalize ballistically, as in the so-called 'quantum hose-race'~\cite{Weitenberg2011}), it allows
for the probing of non-local time-dependent particle Greens functions $\langle \hat{a}_y(t) \hat{a}^{\dagger}_x(0) \rangle$. Specifically, when the system has inversion
symmetry, the sites $x$ and $y$ (coupling respectively to OS 1, 2) are inversion-symmetric and $\beta = i\alpha$ and $\hat{h}_{\rm OS1,2}=0$, we find $\langle \nosopone(t)\rangle_\omega - \langle \nosoptwo(t)\rangle_\omega=\frac{4|\alpha|^2}{\hbar^2}\int_0^t dt_1 \int_0^{t_1} dt_2 \Jperp(t_1)\Jperp(t_2) \operatorname{Im}[\langle\hat{a}_{x-y}(t_1)\hat{a}^\dagger_0(t_2)\rangle]$. In this way, the purely local contributions to $\langle \hat{n}_{\rm OS1,2} \rangle_\omega$ are removed. Second, an atom initially prepared on an OS or two could be given a different internal hyperfine state, making it  distinguishable from the atoms in the system proper. Such schemes would provide a straightforward
way to measure the local and nonlocal Greens function of a mobile distinguishable impurity in 1D or 2D quantum systems, a subject which has attracted considerable theoretical~\cite{Kopp1990,Rosch1995,Zvonarev2007,Kantian2014} and experimental interest recently~\cite{Palzer2009,Catani2012,Koschorreck2012,Kohstall2012,Fukuhara2013}. Third, the cold atoms setup allows one to measure quantities that would be difficult to access in analogous condensed matter experiments, namely fourth-order time-dependent correlators $C_{x_4x_3x_2x_1}(t_4,t_3,t_2,t_1):=\langle \hat{a}^{\dagger}_{x_4}(t_4) \hat{a}^{\dagger}_{x_3}(t_3) \hat{a}_{x_2}(t_2) \hat{a}_{x_1}(t_1) \rangle$.
Measuring the time-evolving density-density correlator on OS 1 and 2, $\langle\nosopone(t)\nosoptwo(t)\rangle_\omega$ (as in \cite{Cheneau2012}), coupling to lattice-sites $x$ and $y$ respectively with the same tunnel coupling, will yield
\begin{widetext}
\begin{eqnarray}\langle\nosopone(t)\nosoptwo(t)\rangle_\omega & = & \frac{1}{\hbar^4}\int_0^t dt_1 \int_0^{t_1} dt_2 \int_0^t dt'_1\int_0^{t'_1} dt'_2 \Jperp(t_1) \Jperp(t_2)  \Jperp(t'_2) \Jperp(t'_1) \nonumber \\ & & \times \left[ C_{xyyx}(t_1,t_2,t'_2,t'_1) + C_{yxxy}(t_1,t_2,t'_2,t'_1) \pm C_{xyxy}(t_1,t_2,t'_2,t'_1) \pm C_{yxyx}(t_1,t_2,t'_2,t'_1) \right], \end{eqnarray} 
\end{widetext}
where $\pm$ is the sign for bosons (fermions). As for eqs.~(\ref{lin_resp_eq_init_empty}), (\ref{lin_resp_eq_init_occ}), we assume only single-component tunneling to and from OS 1, 2. The detailed implications of these three extensions will be the subject of future work.

In conclusion we have demonstrated a schemed to measure correlation functions of ultracold atomic gases confined to optical lattices. This scheme only requires local addressability, which is available in today's experiments. 
We have validated the scheme by comparing it to t-DMRG calculations and proposed extensions to deal with non-local and higher order correlations.

This work was supported in part by the Swiss NSF under division II, and the ARO-MURI grant (W911NF-14-1-0003). US acknowledges funding by the DFG through FOR801 and NIM.

\bibliography{/Users/kantian/Documents/library.bib}

\end{document}